\providecommand{\main}{.}
\documentclass[journal ]{new-aiaa}
\usepackage{amsmath}
\usepackage[utf8]{inputenc}
\usepackage{textcomp}

\usepackage{graphicx}
\graphicspath{{images/}{\main/images/}{images/deflection/}{\main/images/flyby/}{\main/images/loiter/}{\main/images/rendezvous/}{\main/images/sample_return/}}

\usepackage{savesym}
\savesymbol{tablenum}
\usepackage{siunitx}
\restoresymbol{SIX}{tablenum}
\usepackage{longtable,tabularx}
\newcommand{\hlreviewone}[1]{{#1}}

\title{Preliminary Analysis into the Feasibility of Missions to Asteroid 2024 YR$_4$}

\author{Adam Hibberd \footnote{Research and Software Engineer, Initiative for Interstellar Studies (i4is), United Kingdom, adam.hibberd@i4is.org (Corresponding Author)}}
\affil{Initiative for Interstellar Studies (i4is), 27/29 South Lambeth Road London, SW8 1SZ United Kingdom}
\author{T. Marshall Eubanks\footnote{Head Scientist, Space Initiatives Inc., USA}}
\affil{Space Initiatives Inc, Princeton, WV 24740, USA}

\begin{document}

\maketitle

\section*{Nomenclature}

{\renewcommand\arraystretch{1.0}
\noindent\begin{longtable*}{@{}l @{\quad=\quad} l@{}}
$\Delta V$  & Velocity increment (\si{km.s^{-1}}) \\
$V_{\infty}$ &    Hyperbolic excess speed relative to Earth (\si{km.s^{-1}})\\
$C_3$& Characteristic energy of escape orbit from Earth   $ = V_{\infty}^2$
 (\si{km^2.s^{-2}}) \\
$\beta$ & Spacecraft impact momentum enhancement factor (no units)\\
M & Mean anomaly (\si{rad})\\
n & Mean motion (\si{rad.s^{-1}})\\
T, t & Time (\si{s})\\
$\mu$ & Gravitational mass of Sun (\si{m^3s^{-2}})\\
a & Semi-major axis (\si{m})\\
E & Orbital energy per unit mass (\si{m^2s^{-2}})\\
\textbf{V} & 2024 YR$_4$ velocity vector (\si{m.s^{-1}})\\
$\Delta$R & Longitudinal displacement required from impact (\si{m})\\ 
\textbf{V$_i$} &  Velocity of impactor (\si{m.s^{-1}})\\
V$_{long}$ & Component of impactor velocity along target's velocity vector (\si{m.s^{-1}})\\
m$_i$ & Mass of impactor required (\si{kg})\\
m$_{yr4}$ & Mass of 2024 YR$_4$ (\si{kg})\\

\end{longtable*}}

\setcounter{table}{0}


\section{Introduction}
\label{sec1}

Near Earth object (NEO) designated 2024 YR$_4$, which also ranks as a potentially hazardous object (PHO), was discovered by the Asteroid Terrestrial-impact Last Alert System (ATLAS) on $27^{th}$ December 2024 \citep{2024MPEC....Y..140W}. Its initial impact probability was 1 \% and, as is often the case with such discoveries, began to increase with further observations, reaching a high of 3.2 \%, with a correspondingly high Torino impact hazard score, slightly less than that temporarily reached by Apophis (2024 MN$_4$) following its discovery in December 2004. At the time of writing this probability has now significantly diminished to an almost negligible level (0.36 \%), yet missions are still of considerable interest to humanity, especially in view of the recommendation by the 2022 Planetary Decadal Survey \citep{2022cosp...44..403C} which, to quote, states that ''The highest priority planetary defense demonstration mission...should be a rapid-response, flyby reconnaissance mission targeted to a challenging NEO, representative of the population ($\sim{50}-100$ m in diameter) of objects posing the highest probability of a destructive Earth impact''.\\
\hlreviewone{When it comes to missions to NEOs a key achievement was the so-called Halley Armada contemporaneous with the return of Halley to the inner Solar System way back in 1986. Amongst the many flyby probes, the main mission\citep{GIOTTO}, Giotto, was the first to send back images of a comet, as it flew by at a relative velocity of $\sim{69}$ \si{km.s^{-1}}. Subsequent to this groundbreaking achievement, in 2000, the mission NEAR-Shoemaker actually rendezvoused (i.e. matched velocity) with the Near Earth Asteroid (NEA) known as Eros, to enable prolonged exploration of Eros, as it orbited the asteroid \citep{NASA_NEAR_Shoemaker}.\\
The era of sample return was heralded in by the magnificent, though low-key, JAXA mission Hayabusa \citep{JAXA_Hayabusa} in the '00s, which dispatched a lander to pick up actual material from 25143 Itokawa and return it to Earth in 2010 with precious data from the mission. The sample pick-up mechanism was found to have failed but particles from the asteroid were nevertheless still present.\\
Indeed landers on NEOs were to become popular, with for example, the ESA Rosetta mission \citep{ROSETTA} launched in 2004 (Philae) and also the NASA OSIRIS-REx sample return mission to NEA Bennu \citep{OSIRIS_REX}, a sample of which was returned to Earth on 4$^{th}$ September 2023.\\
The more recent NASA DART mission \citep{Richardson_2024}, an attempt to deflect the asteroid Dimorphos, one of a pair in a binary double asteroid system, was a complete success, and results are utilised later in this paper.\\}
In this paper we investigate the short-term (that is within the next 7-8 years) feasibility of spacecraft missions to 2024 YR$_4$. To this ends it employs the preliminary interplanetary mission design tool known as 'OITS' or 'Optimum Interplanetary Trajectory Software' detailed in \cite{OITS_info,AH2} and further expounded in \cite{HIBBERD2021584}. This software permits a user-selection from two possible Non-Linear Problem (NLP) solver options, namely NOMAD \citep{LeDigabel2011} or MIDACO \citep{Schlueter_et_al_2009,Schlueter_Gerdts_2010,Schlueter_et_al_2013}. OITS has been successfully exploited for a range of interplanetary missions with all sorts of different applications, including for example such targets as interstellar objects and terrestrial planets \citep{HPE19,HEL22,HHE20,HH21,HPH21,AH23,HA23}.
In the following analysis the general feasibility of first flyby missions, then sample return missions and finally rendezvous missions are investigated, with a specific mission examined to determine the viability of deflecting the object from a Moon collision, should that be deemed necessary through further refinement of 2024 YR$_4$'s orbit.  

As a side note, we emphasize that OITS adopts a patched-conic assumption in its derivation, and that due to the close proximity of the spacecraft to Earth in the flyby and sample return missions addressed here, there may exist  small errors in the $\Delta$V and hyperbolic excesses calculated by OITS. Be reassured however that such perturbations will be small and could readily be corrected by an appropriate in-flight propulsion system with minimal $\Delta$V requirement.
\hlreviewone{
\section{Method}
 Optimum Interplanetary Trajectory Software (OITS) \citep{OITS_info,AH2,HPH21} is a preliminary interplanetary mission design tool conceived, developed, validated and applied solely by principal author, Adam Hibberd.

 At a rudimentary level it condenses down the complexity of mission design into the problem of finding the optimal encounter times, $t_i~, i=1,2...n$ of the $n$ user-specified sequence of celestial bodies visited on the way to the $n^{th}$ and final target object (in this case 2024 YR$_4$ of course).

To do this, we must know the position/velocity (ephemeris) of these celestial bodies accurately as a function only of time, and this is accomplished by exploiting the NASA JPL SPICE software libraries \citep{NAIF}, with appropriate SPICE binary data kernel files for all bodies included along the way. The reader is referred to the SPICE function 'cspice\_spkezr', responsible for this.

Having so derived the positions and velocities at all the encounter times, we assume the only influence on the probe from one celestial body to the next is the gravitational attraction of the Sun, which allows each trajectory arc in turn to be treated as a Lambert problem, the solution of which has been extensively investigated in the literature (go to \cite{Duan2025-in} for example). The approach adopted by OITS is the so-called 'Universal Variable Formulation' \citep{Bate1971}.

For $n=2$ there are two solutions to the Lambert problem, namely long way 'lw' and short way 'sw', one of which will be prograde, the other retrograde. Thus for $n>2$ user-specifed bodies there are $2^{(n-1)}$ possible permutations whereby the route with minimum overall $\Delta$V$_{tot}$ is selected for optimization by the NLP solver. But how is this $\Delta$V$_{tot}$ calculated?

Clearly we can state in the general case that $\Delta$V$_{tot}$ $= \sum_{i=1}^{k} \Delta V_{i}$. Specifically for flyby of the target, then no change of velocity is implemented at this target, thus $k=n-1$. Alternatively at rendezvous then $k=n$.

OITS is perfectly capable of accommodating multiple gravity assists (GA) of planets, for a detailed explanation as to how this is done go to other literature such as \cite{hibberd2026catching3iatlasusingsolar}.

}
\section{Flyby}
\label{sec2}

\begin{figure}[hbt!]
\centering
\includegraphics[width=1.0\textwidth]{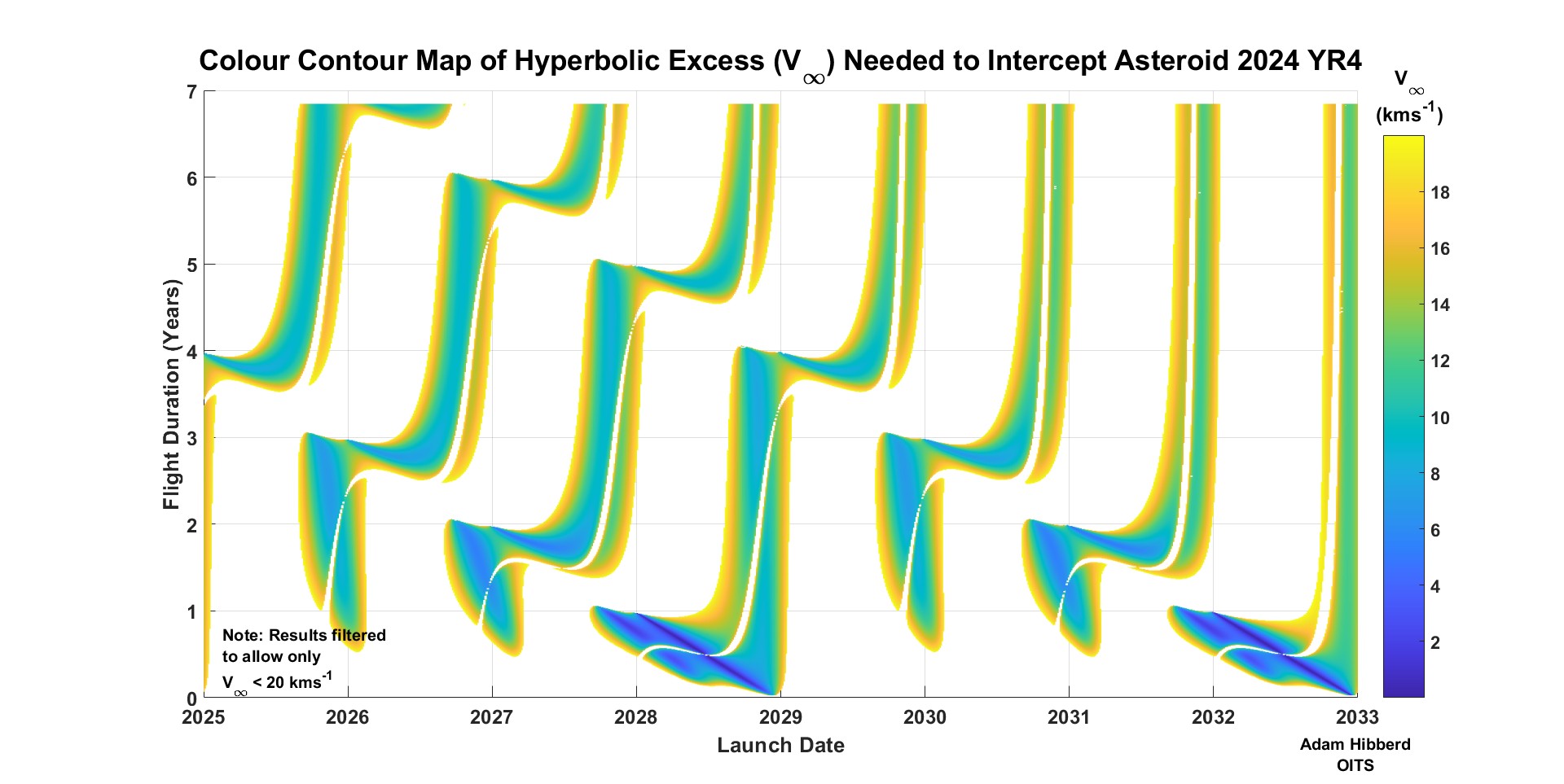}
\caption{Pork Chop plot of flyby missions to 2024 YR$_4$, with colours representing hyperbolic excess (V$_{\infty}$) needed by the launch vehicle to reach the asteroid given launch date (x-axis) and flight duration (y-axis).}
\label{fig:CC1}
\end{figure}

Refer to colour contour plot Figure \ref{fig:CC1}. We find patterns repeating on a 4 year cycle, corresponding to the orbital time period of 2024 YR$_4$. There are two regions of launch feasibility, i.e. from late 2027 throughout most of 2028, and then again on its subsequent passage through perihelion four years later, from late 2031 throughout most of 2032, with resulting launch hyperbolic excess speeds (V$_{\infty}$) on the order of small fractions of a kilometre per second. Optimal flight durations remain largely below 1 year.

The two dark blue bands apparent in both launch windows (i.e. the windows in 2028 and 2032) correspond to the two minimum-energy arrival scenarios, the lower band equivalent to the spacecraft arriving at around the descending node of 2024 YR$_{4}$ (when the object intersects the Ecliptic and also in the case of 2024 YR$_4$ reaches one Earth distance from the Sun) in late December of 2028 and 2032 respectively; and the higher band equivalent to the spacecraft arriving at the point where it reaches exactly one Earth distance from the Sun (i.e. 1 au, yet above the Ecliptic plane) in mid-October of 2028 and 2032 respectively.

\begin{figure}[hbt!]
\centering
\includegraphics[width=1.0\textwidth]{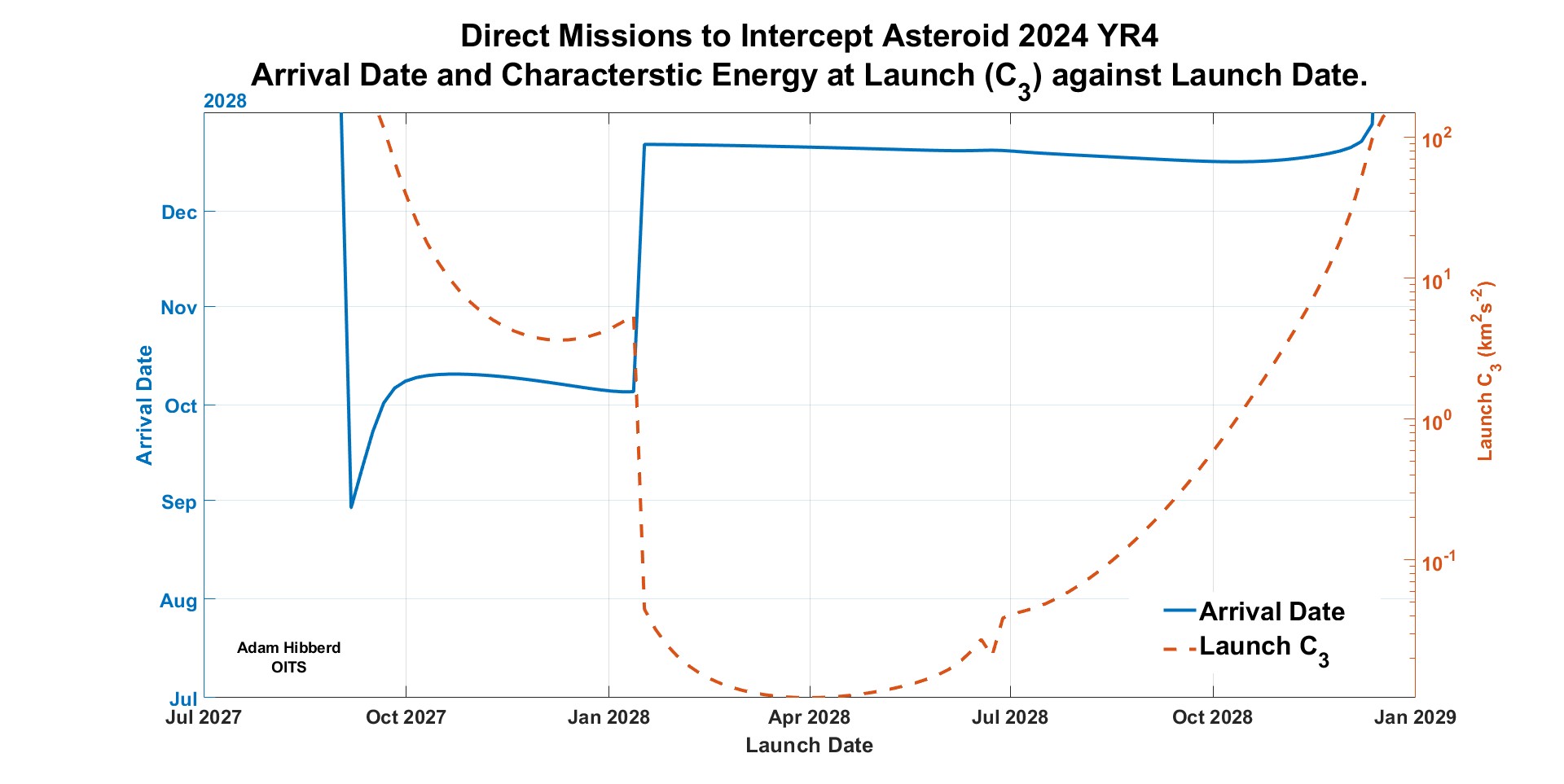}
\caption{A more detailed examination of the optimal 2027-2028 launch windows shown in the Pork Chop plot \ref{fig:CC1}, showing arrival dates - blue solid line, left vertical axis - and Characteristic launch energy (C$_3$) - red dashed line, right vertical axis.}
\label{fig:Plot1}
\end{figure}

\begin{figure}[hbt!]
\centering
\includegraphics[width=1.0\textwidth]{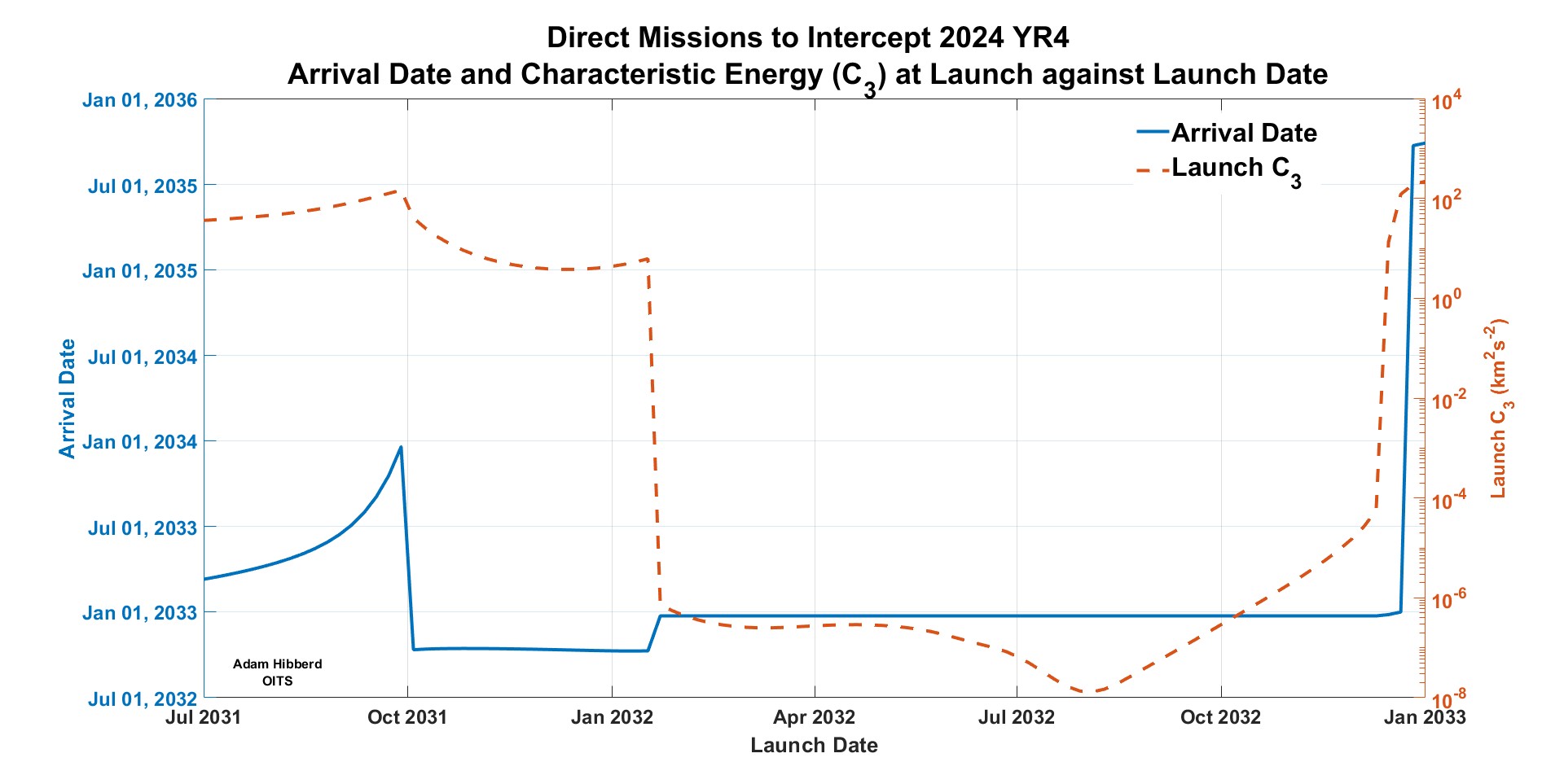}
\caption{A more detailed examination of the optimal 2031-2032 launch windows shown in the Pork Chop plot \ref{fig:CC1}}
\label{fig:Plot2}
\end{figure}

To address each of these 2028 and 2032 flyby launch opportunities in greater detail, refer to Figures \ref{fig:Plot1} and \ref{fig:Plot2} respectively. Observe that the characteristic energy (C$_3 = V_{\infty}^2$) in the 2028 case becomes negligibly small (with a minimum in late March 2028), indicating an almost parabolic Earth escape orbit, readily achievable by a range of currently available launch vehicles, including for example the European Ariane 6\footnote{Note that the predecessor Ariane 5 launched the James Webb Space Telescope into an orbit with a C$_3$ near 0 \si{km^2.s^{-2}}, with extreme accuracy}. Furthermore a spacecraft mission to this object could exploit a ride-share on a launcher bound for the Moon, with very little extra $\Delta$V capability needed by the spacecraft.

\begin{figure}[hbt!]
\centering
\includegraphics[width=1.0\textwidth]{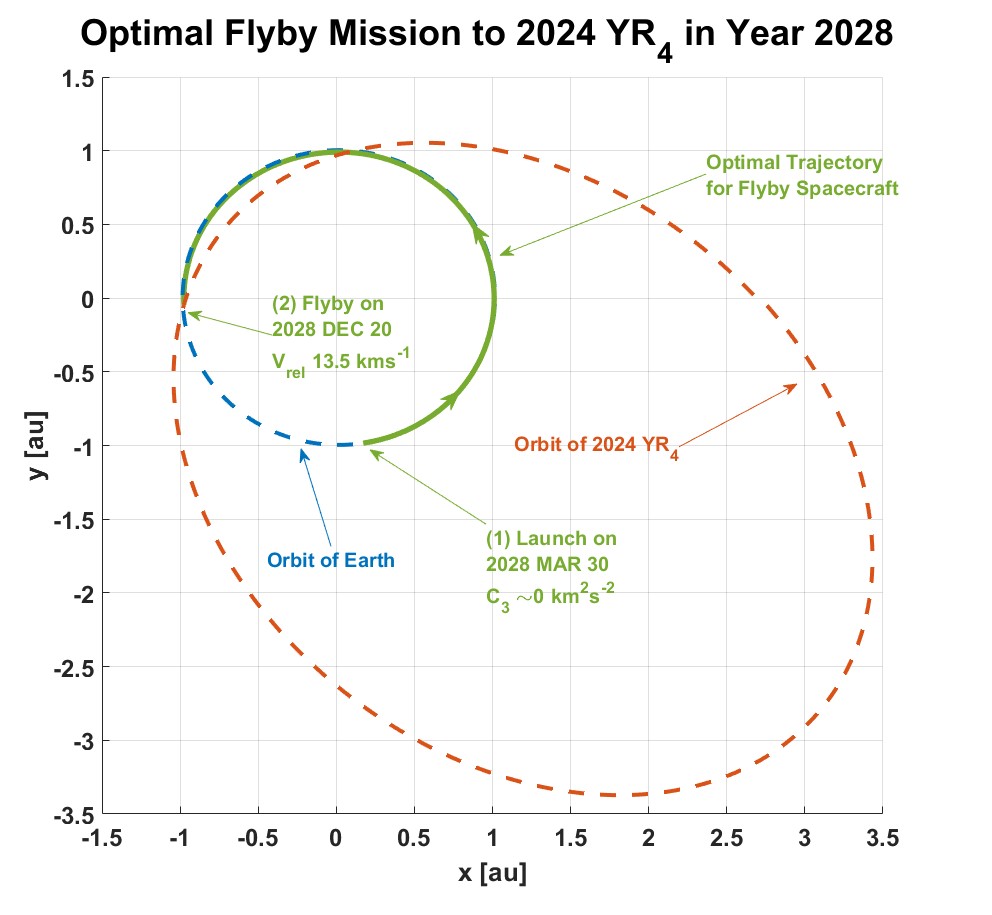}
\caption{Flyby trajectory with minimum Characteristic launch energy (C$_3$) in year 2028}
\label{fig:Plottraj1}
\end{figure}

\begin{figure}[hbt!]
\centering
\includegraphics[width=1.0\textwidth]{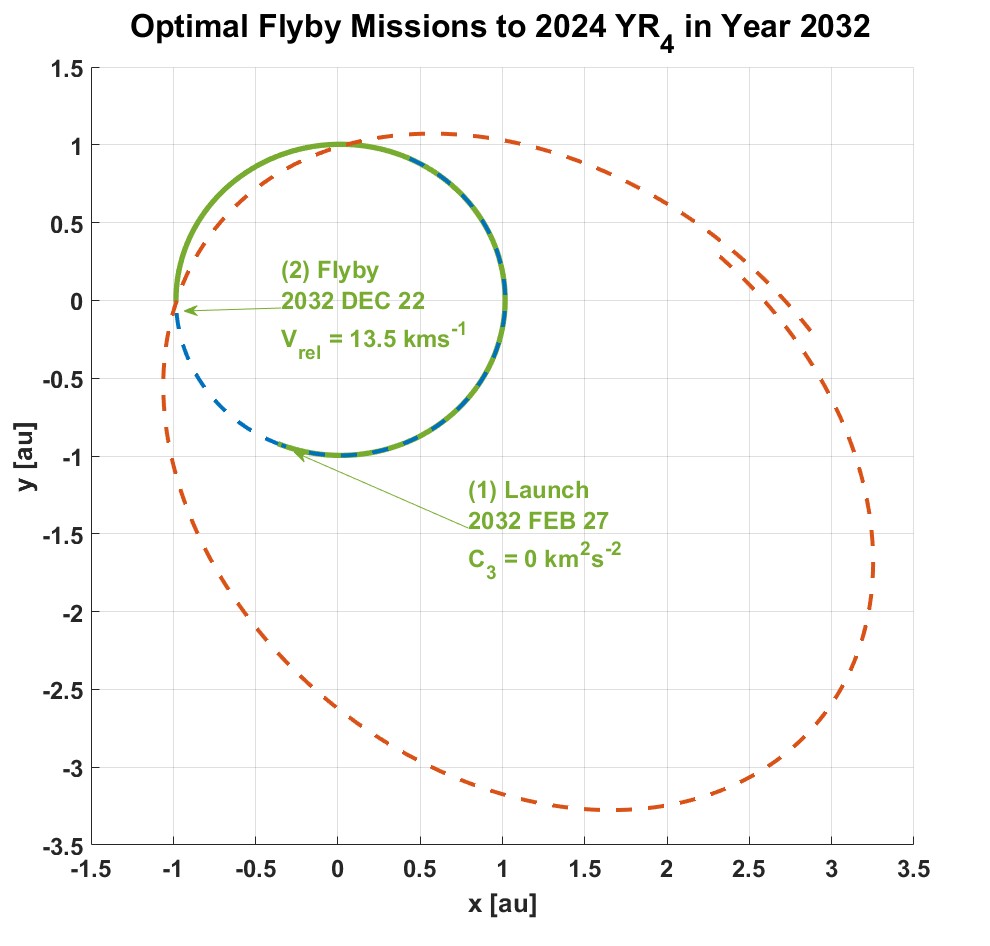}
\caption{Flyby trajectory with minimum Characteristic launch energy (C$_3$) in year 2032}
\label{fig:Plottraj2}
\end{figure}

\hlreviewone{Trajectory plots for the optimal flyby missions arising in launch years 2028 and 2032 are provided in Figures \ref{fig:Plottraj1} and \ref{fig:Plottraj2} respectively. One can see from these plots that from an overhead point-of-view, the path of 2024 YR$_4$ intersects the orbit of Earth at two points. The optimal solution is to arrive at the second opportunity in December, where the displacement from the ecliptic is smaller, but is there any merit in arriving at the first, i.e. in October?}

\begin{figure}[hbt!]
\centering
\includegraphics[width=1.0\textwidth]{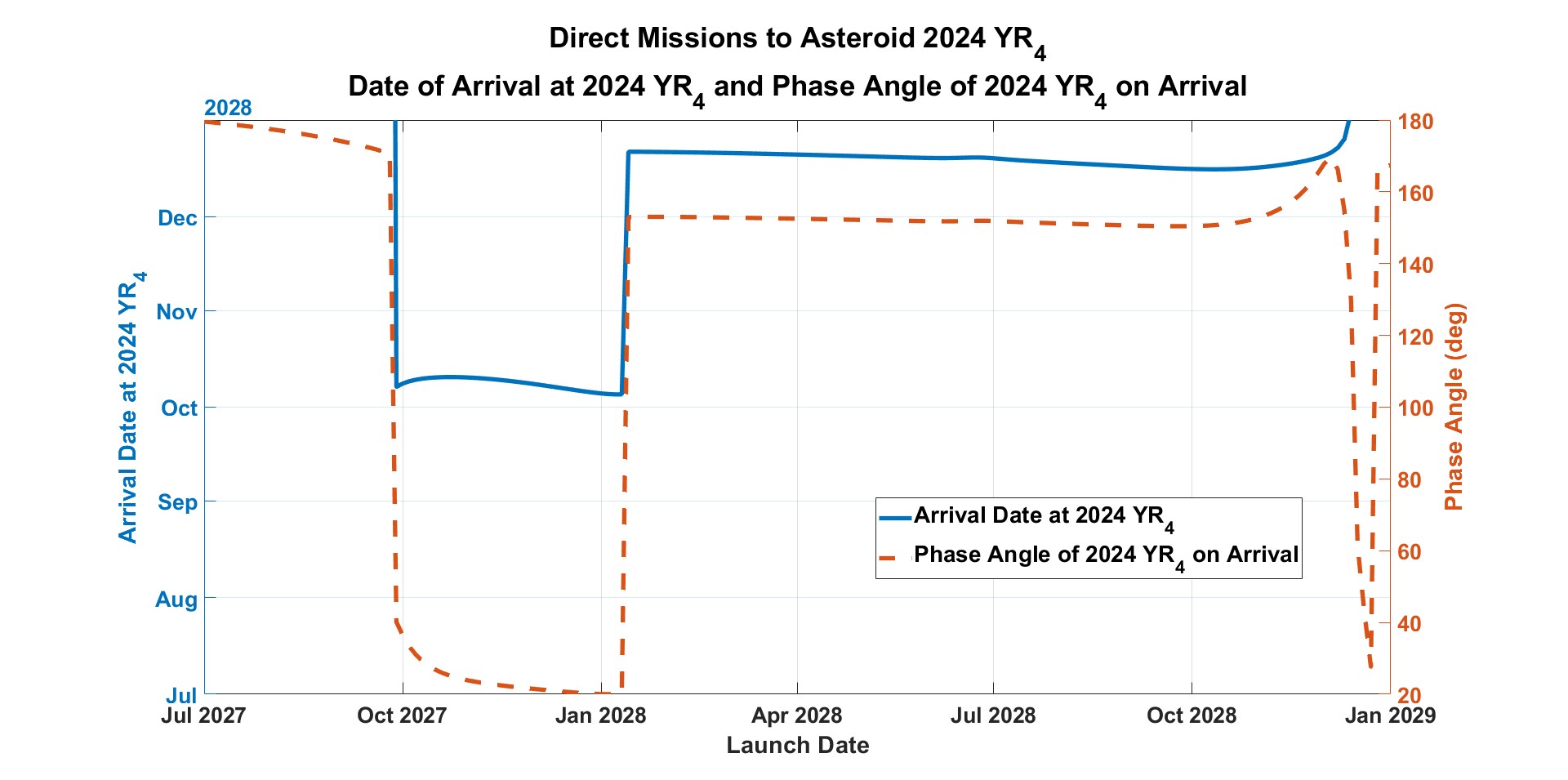}
\caption{Optimal arrival date (with minimum launch C$_3$) and the phase of 2024 YR$_4$ as observed by spacecraft at encounter against launch date from July 2027 to Jan 2029}
\label{fig:PlotPhase}
\end{figure}

\hlreviewone{In an attempt to answer this question, refer to Figure \ref{fig:PlotPhase}. This provides the phase of 2024 YR$_4$ as observed by the probe at arrival (right hand axis) with the optimal arrival date (left axis). This metric gives a vital indicator as to the lighting conditions of the target which in turn impacts on the quality of scientific return garnered from the mission.}

\hlreviewone{We find that for launch dates in 2027 into 2028, there is a launch window opening in October 2027 and closing January 2028, where the optimal arrival date lies precisely in early October, corresponding to the first intersection 2024 YR$_4$ makes with Earth's orbit. Observe the phase is quite low over this period of time ($\sim{20}$ $^{\circ}$) corresponding to excellent lighting conditions for observations, whereas outside of this narrow window, the phase angle is $\sim{160}$ $^{\circ}$ or more, indicating the object is largely silhouetted against the Sun for these particular missions. The conclusion from this is that the minimum C$_3$ missions in 2028 and 2032, although convenient for a launch, are not actually the preferred choice scientifically speaking, and an arrival in early October offers a significant benefit in these terms.}

\begin{figure}[hbt!]
\centering
\includegraphics[width=1.0\textwidth]{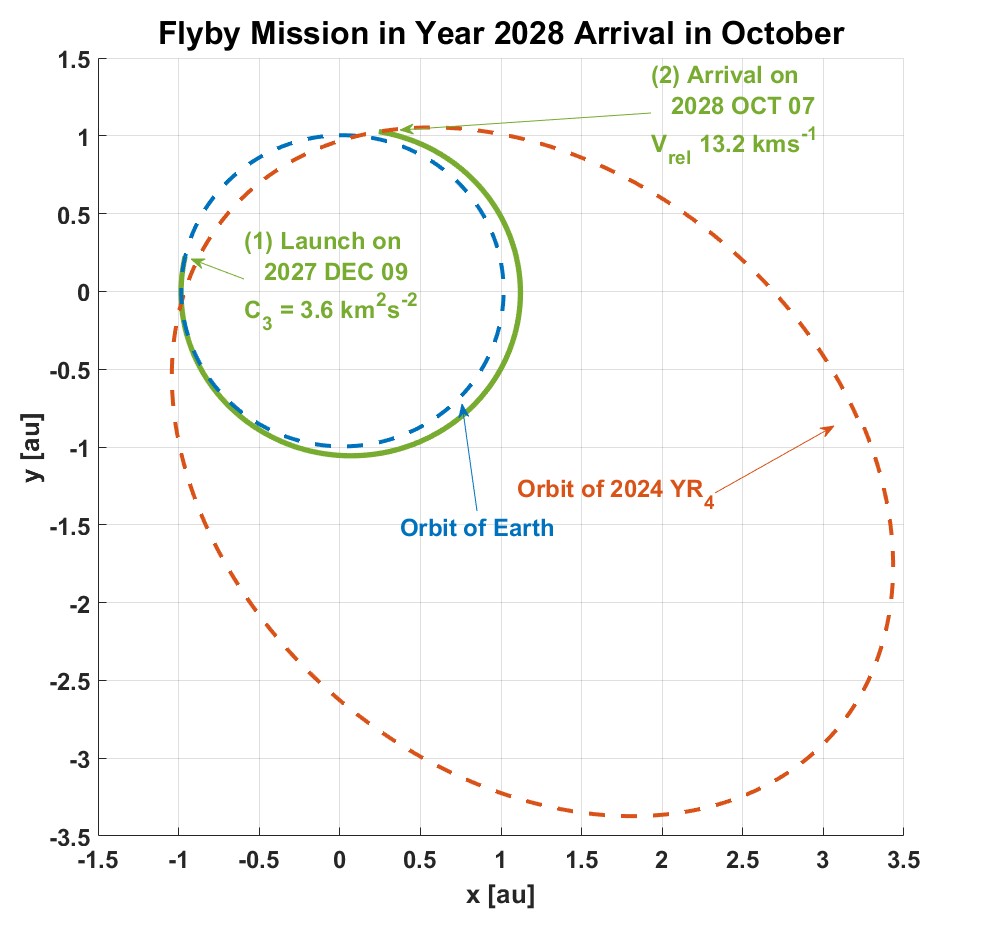}
\caption{Flyby trajectory with arrival in October of year 2028}
\label{fig:Plottraj3}
\end{figure}

\begin{figure}[hbt!]
\centering
\includegraphics[width=1.0\textwidth]{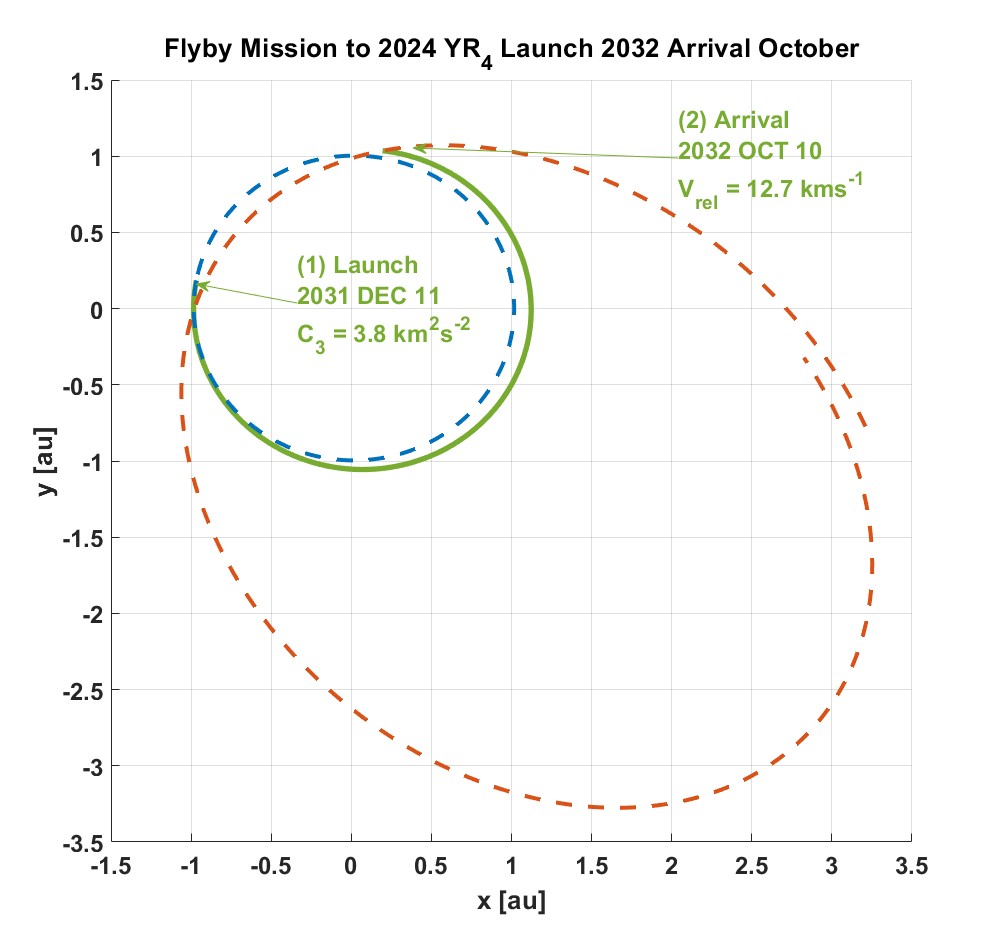}
\caption{Flyby trajectory with arrival in October of year 2032}
\label{fig:Plottraj4}
\end{figure}

\hlreviewone{The mission trajectories for arrival in October of years 2028 and 2032, corresponding to the above-mentioned preferred option, are shown in Figures \ref{fig:Plottraj3} and \ref{fig:Plottraj4} respectively.}

\begin{figure}[hbt!]
\centering
\includegraphics[width=1.0\textwidth]{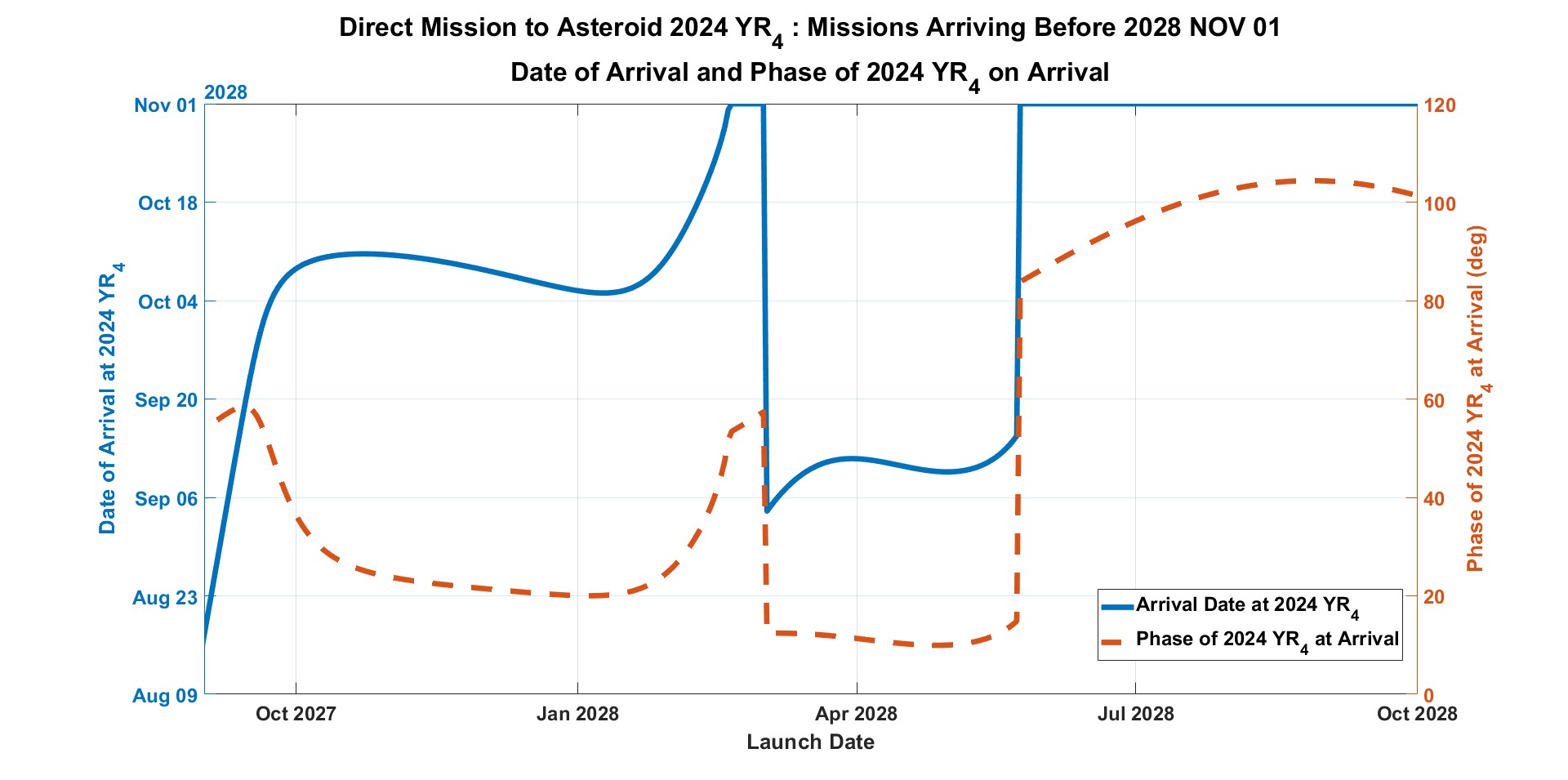}
\caption{Direct Missions in 2028 arriving before November and so with improved lighting conditions: time of arrival (left axis) and phase of target (right axis)}
\label{fig:Phase_Tcons}
\end{figure}

\begin{figure}[hbt!]
\centering
\includegraphics[width=1.0\textwidth]{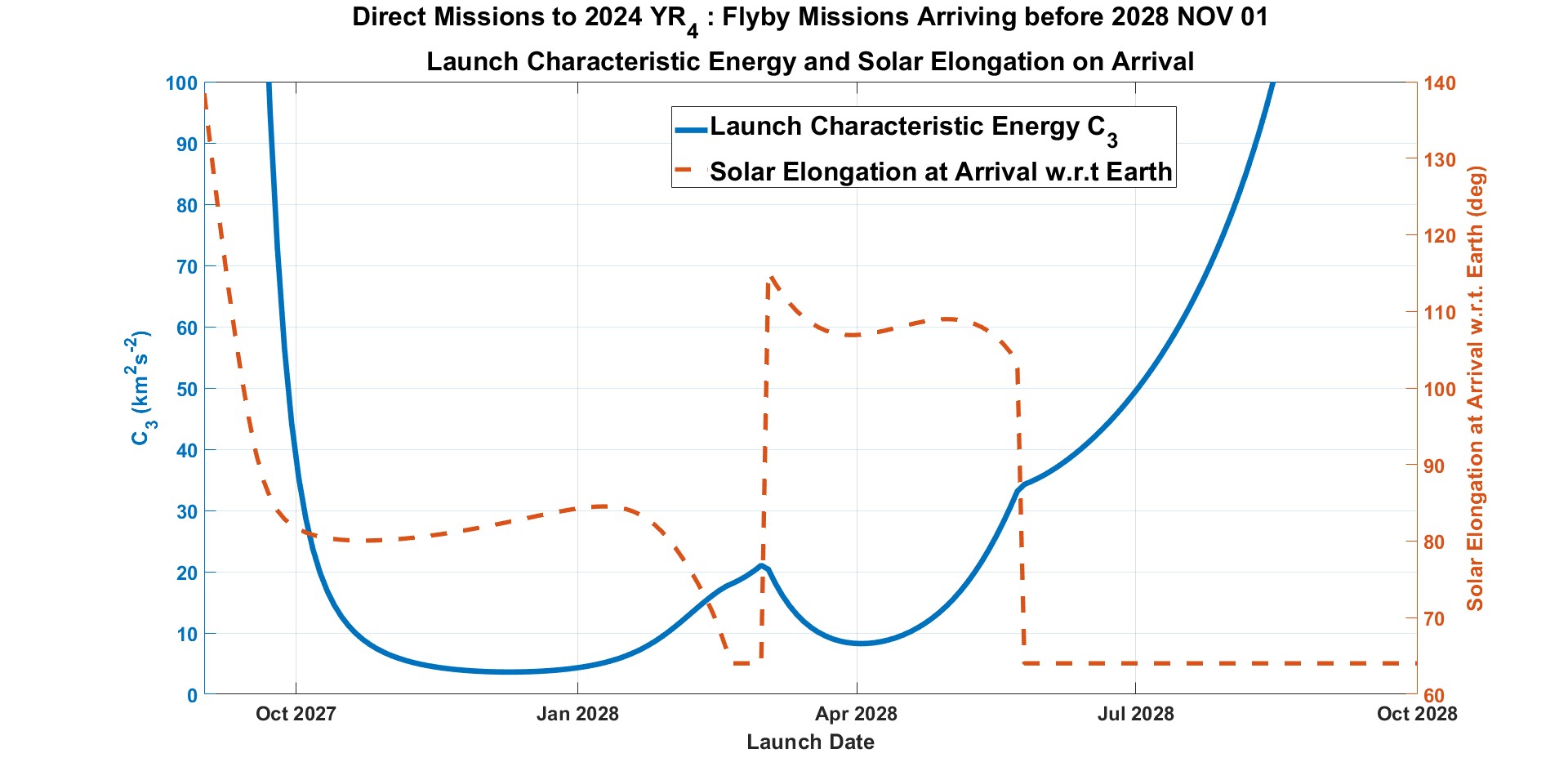}
\caption{Direct Missions in 2028 arriving before November and so with improved lighting conditions: launch C$_3$ (left axis) and solar elongation of target (right axis)}
\label{fig:Phase_Tcons2}
\end{figure}

\hlreviewone{Let us now proceed on the basis that we wish to arrive in October for reasons which have been articulated. In this case refer to Figure \ref{fig:Phase_Tcons} and \ref{fig:Phase_Tcons2}. These impose an arrival constraint on the mission to be earlier than 2028 November. What we find when we examine this, is that the phase is indeed low ( $\sim{20}$ $^{\circ}$) for launches earlier than around the beginning of March 2028, yet for a window beginning at this latter time and closing at the end of May, the arrival dates converge to times in September of 2028, and have an even lower phase (of $\sim{10}$ $^{\circ}$).}

\hlreviewone{Thus this analysis with this arrival date inequality constraint has exposed further feasible launches to 2024 YR$_4$ which otherwise would have remained hidden. Note that Figure \ref{fig:Phase_Tcons2} also provides the launch C$_3$ needed to arrive in September 2028 as having a minimum of $\sim{8}$ $\si{km^2.s^{-2}}$. This value corresponds to around that necessary for a Mars transfer, yet a Mars rideshare is probably not possible since although 2028 is a favourable year for a missions to Mars, the windows happen to lie later on in that year.}

\hlreviewone{To summarise, the optimal direct routes to flyby 2024 YR$_4$, arriving in December 2028/2032, offer attractively low launch requirements, but are lacking considerably since the object is poorly lit by the Sun w.r.t to the arriving probe. Instead arrival dates in October 2028/2032 need only marginally higher launcher performance, yet the asteroid is just about completely exposed to the Sun from the probe's viewpoint which is ideal for science return.}

\subsection{Moon-Avoidance Kinetic Deflection Mission}

As of the time of writing, the probability of 2024 YR$_4$ striking the Moon has \hlreviewone{ dropped to a negligible level, though has in the past been as high as 4\%. \citep{MOONHIT} We can however treat 2024 YR$_4$ as a test case for a  a kinetic deflection mission to divert the object based on the assumption that otherwise there would be a highly energetic impact with the Moon (around $\sim{6}$ megatonnes of TNT). This is derived adopting an impact velocity of 13.49 \si{km.s^{-1}} and a mass of 2.9 $\times 10^8$ \si{kg}. The latter assumes a diameter of 60 \si{m} \citep{Rivkin_2025} and a density of 2.6 \si{g.cm^{-3}} \citep{SENTRYYR4}, with the standard equation for kinetic energy. Let us proceed on this basis.}

\begin{figure}[hbt!]
\centering
\includegraphics[width=1.0\textwidth]{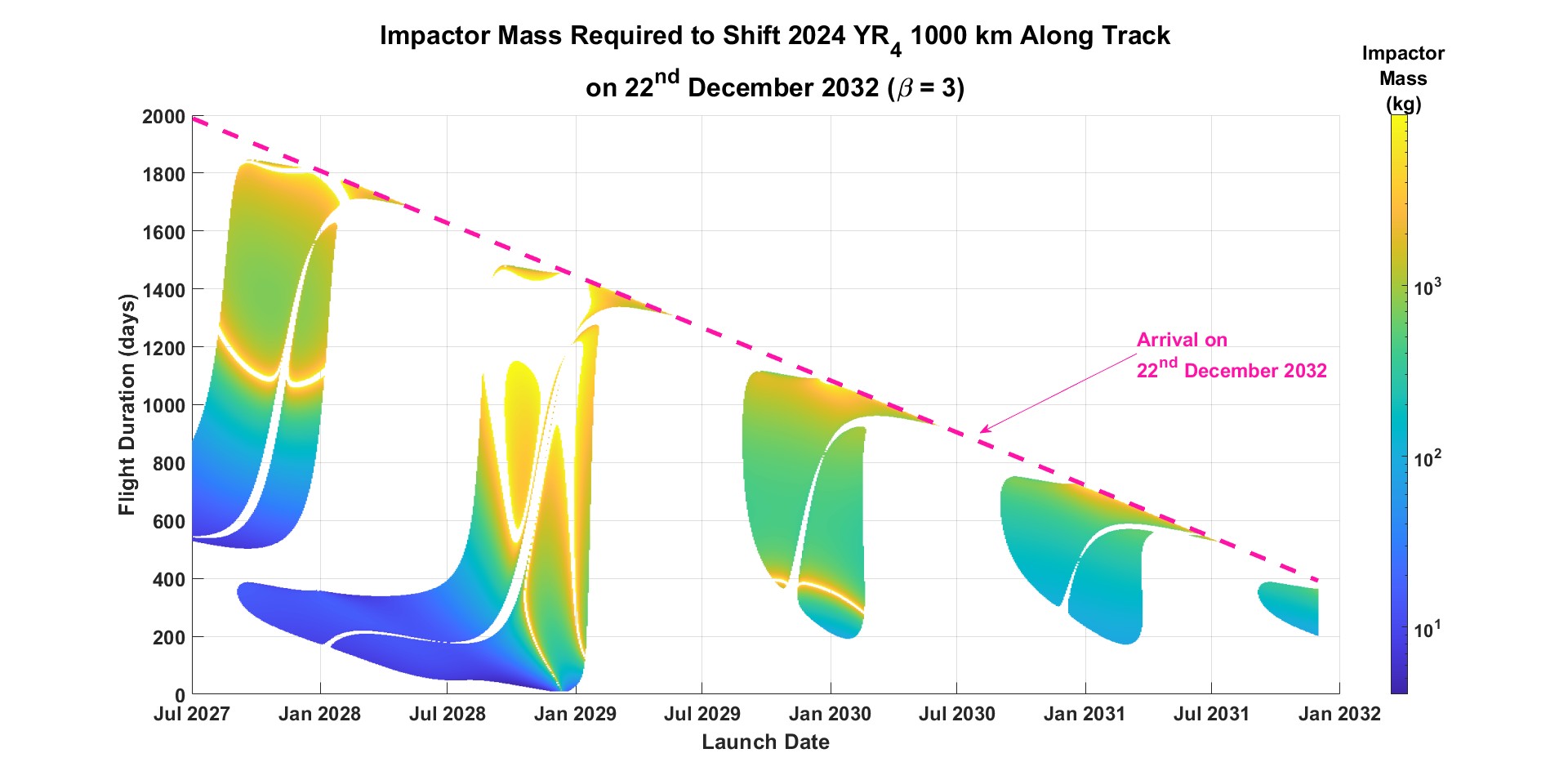}
\caption{Colour Contour Map of impactor masses necessary to deflect 2024 YR$_4$ and shift it by 1000 \si{km} along track so as to avoid either Earth or, as is considered in this paper the Moon, on $22^{nd}$ December 2032}
\label{fig:Plot2.5}
\end{figure}

\hlreviewone{When considering the dynamics of an impactor, in particular the optimal direction for it to deliver an impulse to the target so as to deflect it as far away as possible from an otherwise inevitable collision with the Moon, two approaches present themselves, first a push in the target's direction of motion, and second a push in some direction perpendicular - or normal - to this.

In principle, a suitably chosen normal push, would result in a corresponding displacement of the target away from the Moon, but such a push would NOT affect significantly the energy of the object, meaning its time period will remain unchanged, leading to the problem of regular encounters further down the line, representing a continuing hazard.

Not so a longitudinal (or tangential) push in the direction of the target's motion. Note that his will also displace the target to a certain degree in position, but in addition and more importantly will also significantly alter its orbital energy. This has a direct impact on its 'mean motion', $n$ (otherwise known as its mean angular velocity), and consequently its time period. Thus there will be a cumulative positional error with passing time, making for a far more effectual - and increasing - level of deflection over time. The following lays out the astrodynamics of an impact intended to apply a $\Delta$V in the direction of motion. 

First of all the Mean Anomaly, $M$ is expressed in terms of mean motion as follows:
\begin{equation}
\label{MA}
    M = n\left(T-t\right)
\end{equation}
where T is the future time of collision with the Moon and t is the time of impulse delivered by the impactor. 

As mentioned, a $\Delta$V applied to the target in its direction of motion $V$, will directly affect the mean motion, $n$, but how?

Well $n$ depends only upon the semi-major axis, $a$, of the asteroid's orbit, thus:
\begin{equation}
\label{mm}
    n = \sqrt{\frac{\mu}{a^3}}
\end{equation}
$\mu$ being the gravitational mass of the Sun. Continuing on from this, we also know that $a$ is dependent only on the orbital energy, $E$, as follows:
\begin{equation}
\label{sma}
    a = -\frac{\mu}{2E}
\end{equation}
The kick in the asteroid's velocity $\Delta$V, will change the kinetic energy by:
\begin{equation}
    \Delta E = V \Delta V
\end{equation}
which is from the well-known expression for kinetic energy.
Differentiating equation \ref{sma} yields:
\begin{equation}
    \frac{\Delta a}{a} = -\frac{2a}{\mu}V\Delta V
\end{equation}
If we now differentiate $n$ w.r.t. $a$, as given in equation \ref{mm}, we find:
\begin{equation}
\label{deln}
    \Delta n = -\frac{3}{2}n\left(\frac{\Delta a}{a}\right) =  3n\frac{a}{\mu}V\Delta V= 3V \Delta V\sqrt{\frac{1}{\mu a}}
\end{equation}
From \ref{MA} ans \ref{deln} above, clearly:
\begin{equation}
    \Delta M =\Delta n \left(T-t\right) = 3V \Delta V\left(T-t\right)\sqrt{\frac{1}{\mu a}}
\end{equation}

Now let's say we need to shift the asteroid longitudinally by distance $\Delta$R, this requires a change in mean anomaly $M$ of:
\begin{equation}
    \Delta M = \frac{\Delta R}{a}
\end{equation}

Since to first order $\Delta V = \Delta M/\left(\frac{dM}{dV}\right)$, we can see that the required change in velocity longitudinally would be:
\begin{equation}
\label{dv}
  \Delta V = \frac{\Delta R \sqrt{ua}}{3aV\left(T-t\right)}  =  \frac{\Delta R }{3V\left(T-t\right)}\sqrt{\frac{\mu}{a}}
\end{equation}
But how do we relate this required change in longitudinal velocity to required impactor mass, $m_i$?

Let us say that the impactor is travelling relative to 2024 YR$_4$ at encounter with a velocity vector \textbf{V$_i$}. Note the bold font now indicates a vector. Furthermore, we suppose that 2024 YR$_4$ has a heliocentric velocity vector given by \textbf{V}. The longitudinal component of the impactor's velocity, V$_{long}$ is thus:
\begin{equation}
\label{vlong}
    V_{long} = \frac{\mathbf{V_i}\cdot \mathbf{V}}{V}
\end{equation}
where the dot indicates dot product and of course $V=\left|\mathbf{V}\right|$. This impact velocity V$_{long}$ will change the longitudinal velocity of 2024 YR$_4$ through the following relationship:
\begin{equation}
\label{beta}
    \Delta V = \beta V_{long} \frac{m_i}{m_{yr4}}
\end{equation}

In equation \ref{beta}, the parameter $\beta$ is known as the impact ratio (more on this later), and m$_{yr4}$ is the mass of 2024 YR$_4$. Rearranging:
\begin{equation}
    m_i = \frac{m_{yr4}}{\beta}\frac{\Delta V}{V_{long}}
\end{equation}
We can now replace $\Delta$V by \ref{dv} and V$_{long}$ by \ref{vlong}, whence we find:
\begin{equation}
    m_i = \frac{m_{yr4}\Delta R }{3\beta\left(T-t\right)\left(\mathbf{V_i}\cdot \mathbf{V}\right)}\sqrt{\frac{\mu}{a}}
\end{equation}

}
Refer now to Figure \ref{fig:Plot2.5} which provides the mass of impactor necessary to shift the target 2024 YR$_4$ by an arbitrarily chosen distance of \hlreviewone{$\Delta R = 1000 \si{km}$} when it reaches its perilune distance on $22^{nd}$ December (this perilune would clearly be less than one Moon radius should a collision be predicted). To generate this plot, reference parameters are drawn from the successful DART mission, refer to \cite{Richardson_2024}. In particular a reference impact ratio $\beta$ of 3 is adopted, within the range determined post-impact for the DART mission. In addition, a mass for 2024 YR$_{4}$ of \hlreviewone{$m_{yr4} =  2.9 \times 10^8$ kg} is supposed (extracted from \cite{SENTRYYR4}).

We find in this contour map that for launches in late 2027 and for most of 2028, masses on the order of only tens to hundreds of kg would be quite sufficient to shift 2024 YR$_4$ by 1000 km when the object returns to the Earth/Moon system in 2032. Note this is on the order of a Moon radius ($\sim{1700}$ \si{km}) and so provides an excellent estimate of the size of impactor required in order that any Moon collision may be successfully avoided.

\label{sec2_2}

\section{Sample Return with Impactor}
\label{sec3}
We now examine the possibility of sample return missions from 2024 YR$_4$. To summarise these are mission architecures where the spacecraft returns back to Earth after encountering the target, with a cargo of material from it. This sample might be gathered by flying through a tail in the case of a comet as target (like in the NASA Stardust mission, \cite{STARDUST}), or alternatively by dispatching an impactor and flying through the resulting plume in the case of an asteroid.

\hlreviewone{The main astrodynamical requiements of sample return by impactor are as follows: 
\begin{enumerate}
    \item finding a path to the target asteroid and a return path from it which ideally requires no significant in-flight $\Delta$V, AND
    \item to fly through the impact plume with a velocity low enough for the sample retrieval to be performed with optimal efficacy, without deterioration of the captured plume particulates in any way. For aeorgel this means a relative velocity, V$_{rel}$ $\lesssim{6}$ $\si{km.s^{-1}}$.
\end{enumerate} 
}

\hlreviewone{For the moment we shall address the first of these, and it} so happens that if the time period of the spacecraft's outbound orbit towards its target is exactly equally to 1 year \hlreviewone{(or for that matter some integer multiple of 1 year)}, then as well as encountering the target at some point during that year, it will also eventually return to the Earth's orbit, and moreover the Earth will be conveniently located there, since clearly by definition its time period is also a year.

\begin{figure}[hbt!]
\centering
\includegraphics[width=1.0\textwidth]{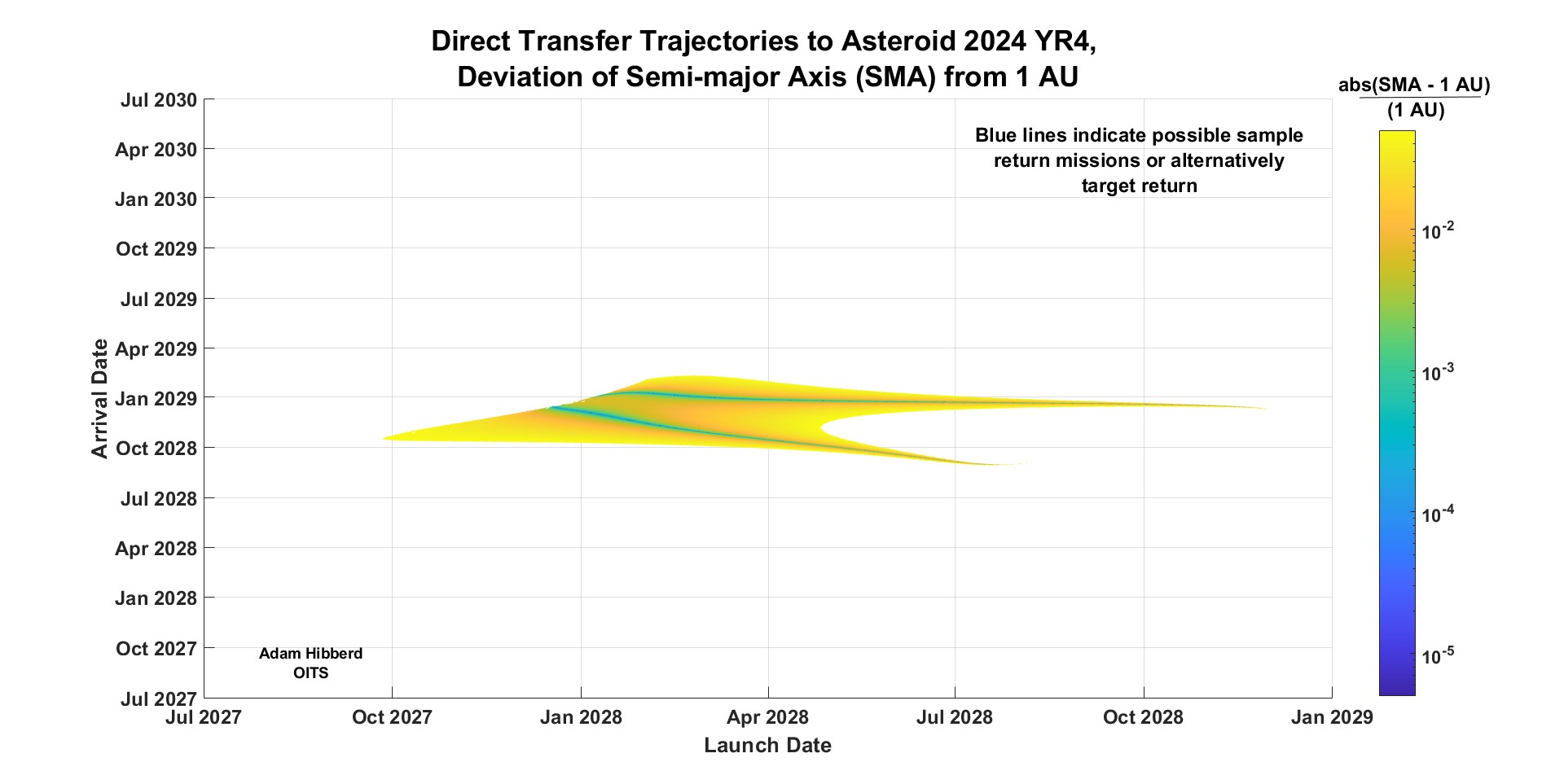}
\caption{Blue bands are bands of opportunity for sample return missions, i.e. launch dates and flight durations the spacecraft might exploit to return back to Earth after exactly one Earth year.}
\label{fig:Plot3}
\end{figure}

There is a direct relationship between semi-major axis and time period of an orbit, meaning that if two orbits have the same time period, they will also necessarily have the same semi-major axis. Now refer to Figure \ref{fig:Plot3}. This plot indicates, for a launch in 2028, the degree of discrepancy between the semi-major axis of the interplanetary orbit taken by any spacecraft mission bound for 2024 YR$_4$, and that of Earth's orbit (1 au) so, by the preceding logic, highlights the possible sample return scenarios. 

Yet again we find that two bands are clearly visible on this plot, this time indicating viable sample return opportunities with arrival at the target either in October of 2028, or later in December 2028, corresponding to 2024 YR$_4$ reaching respectively its pre-perihelion orbital radius of 1 au and then the equivalent point post-perihelion where, as already mentioned, it additionally intersects the Ecliptic plane.

To get a good idea of the $V_{\infty}$ required for these sample return missions refer back to Section \ref{sec2} and Figure \ref{fig:CC1}, we find an extremely low required launch vehicle capability, as would be expected of such a mission.  

\hlreviewone{Further to look into the second of the aforementioned basic requirements, that of arriving with V$_{rel} \lesssim{6} \si{km.s^{-1}}$, we can with a small adjustment to the OITS code, alter the objective function from one of minimizing overall $\Delta$V to one of minimizing V$_{rel}$. Also an additional constraint to enforce NO $\Delta$V at the encounter, is imposed to satisfy the requirement (1) above.}

\begin{figure}[hbt!]
\centering
\includegraphics[width=1.0\textwidth]{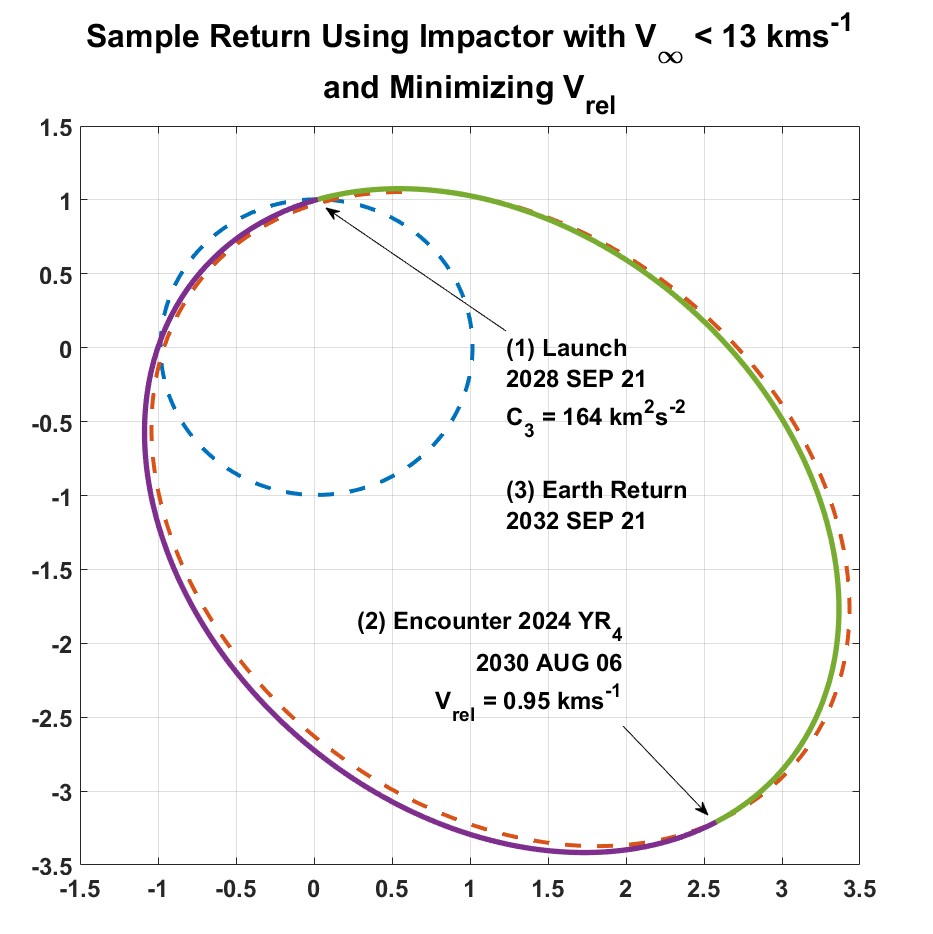}
\caption{Trajectory for sample return with impactor, assuming a C$_3$ less than New Horizons of $\sim{13}$ $\si{km^2.s^{-2}}$ and minimizing encounter velocity, V$_{rel}$}
\label{fig:Plotsri}
\end{figure}

\hlreviewone{Moreover we constrain V$_{\infty}$ at Earth launch to 13 $\si{km.s^{-1}}$ (C$_3$ $=$ 169 $\si{km^2.s^{-2}}$), equivalent to the New Horizons mission \citep{weaver2008new}. This reference is chosen since it is a mission demonstrated from concept to implementation with an extreme C$_3$ compared to other interplanetary missions, in other words it is ideal for feasibility studies. The resulting trajectory is depicted in Figure \ref{fig:Plotsri}.

Observe that the total available V$_{\infty}$ $=$ 13 $\si{km.s^{-1}}$ is not exploited but instead the optimal value turns out to be lower at $\sim{12.8}$ $\si{km.s^{-1}}$ (equivalent to a C$_3$ of 164 $\si{km^2.s^{-2}}$). The encounter velocity is 0.95 $\si{km.s^{-1}}$ and so well under the preferred limit for aerogel of 6 $\si{km.s^{-1}}$}
\hlreviewone{
\section{Sample Return by Rendezvous/Loiter/Return}
\label{secV}
In this section we deal with missions which rendezvous with the target and then loiter for close-up study of the asteroid in situ. A lander is dispatched over this period  to pick up a sample of the asteroid's material, which is then returned back to the waiting spacecraft bus. From here the spacecraft will apply another $\Delta$V, of around $\sim{200}$ $\si{m.s^{-1}}$ to set it on a course back to Earth. The questions are how to reduce the rendezvous $\Delta$V to as low a level as possible, and also how to reduce the launch C$_3$ so that a launcher can inject a sufficient mass to Earth escape? This mass must be large enough to accommodate the rendezvous propellant and also the lander, together with all the necessary instrumentation. The answer, in the case of the historical OSIRIS-REx mission for example, was to conduct a sequence of Earth GAs and Deep Space Manoeuvres (DSMs) so that the orbit of the spacecraft could be gradually matched to that of the target asteroid (in that case asteroid 'Bennu').

Let us adopt the OSIRIS-REx probe, which had a launch mass of 2110 $kg$ and an in-flight $\Delta$V performance of $\sim{1.5}$ $\si{km.s^{-1}}$, as a model for our rendezvous mission to 2024 YR$_4$. The question is ``is such a mission feasible to 2024 YR$_4$?''

A search using OITS was initiated to this effect, and the result is summarized in Figure \ref{fig:Plotsrl}, with Table \ref{tab:Osiris_can_do} providing the mission's vital statistics. The solar elongation of 2024 YR$_4$ with respect to the Earth at the time of rendezvous turns out to be $\sim{156}^{\circ}$, indicating beneficial arrival conditions for communications with Earth. Refer to \cite{VideoAH} for an animation of this mission.

\begin{figure}[hbt!]
\centering
\includegraphics[width=1.0\textwidth]{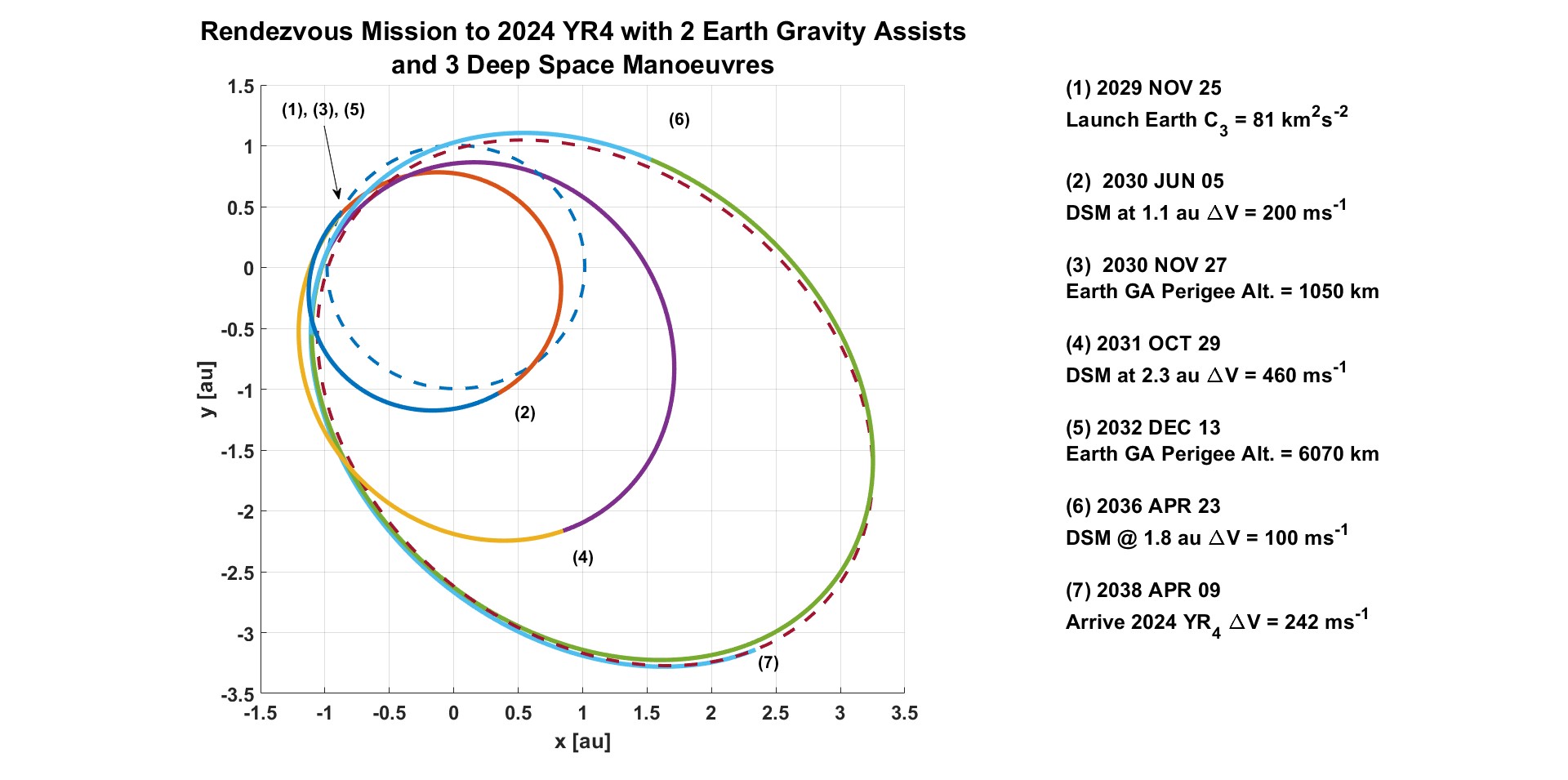}
\caption{Trajectory for sample return with loiter, in the style of the OSIRIS-REx mission to asteroid 'Bennu'}
\label{fig:Plotsrl}
\end{figure}
}

\hlreviewone{
\begin{table}[]
\centering
\begin{tabular}{cccccccc}
\hline
\textbf{} & \textbf{}          & \textbf{}     & \textbf{Arrival} & \textbf{Depart.} & \textbf{}          & \textbf{Cum.}      & \textbf{Periapsis} \\
\textbf{} & \textbf{Encounter} & \textbf{Date} & \textbf{Speed}   & \textbf{Speed}   & \textbf{$\Delta$V} & \textbf{$\Delta$V} & \textbf{Alt.}      \\
\textbf{} & \textbf{}          & \textbf{}     & \textbf{$\si{km.s^{-1}}$}    & \textbf{$\si{km.s^{-1}}$}    & \textbf{$\si{km.s^{-1}}$}      & \textbf{$\si{km.s^{-1}}$}      & \textbf{km}        \\ \hline
\textbf{1} & Earth        & 2029 NOV 25 & 0       & 8.9997  & 8.9997 & 8.9997  & N/A    \\ \hline
\textbf{2} & DSM @ 1.1 au & 2030 JUN 05 & 27.0715 & 27.0191 & 0.2    & 9.1997  & N/A    \\ \hline
\textbf{3} & Earth GA     & 2030 NOV 27 & 9.0526  & 9.0678  & 0.01   & 9.2097  & 1050.4 \\ \hline
\textbf{4} & DSM @ 2.3 au & 2031 OCT 29 & 14.6542 & 14.1952 & 0.46   & 9.6697  & N/A    \\ \hline
\textbf{5} & Earth GA     & 2032 DEC 13 & 11.5132 & 11.5254 & 0.01   & 9.6797  & 6070.1 \\ \hline
\textbf{6} & DSM @ 1.8 au & 2036 APR 23 & 25.2215 & 25.3086 & 0.0999 & 9.7795  & N/A    \\ \hline
\textbf{7} & 2024 YR4     & 2038 APR 09 & 0.2418  & 0       & 0.2418 & 10.0214 & N/A    \\ \hline
\end{tabular}

\caption{Possible rendezvous/loiter/sample return mission to 2024 YR$_4$, where the speeds quoted are relative to the encounter object in question, and the initial $\Delta$V at Earth is the hyperbolic excess speed, V$_{\infty}$}
\label{tab:Osiris_can_do}
\end{table}

\begin{table}[]
\centering
\begin{tabular}{lcccccccc}
\hline
 & \textbf{}          &  & \textbf{Arrival} & \textbf{Depart.} & \textbf{}      & \textbf{Cum.}  & \textbf{Eclip.} & \textbf{Eclip.} \\
 & \textbf{Encounter} & \textbf{Time}    & \textbf{Speed}   & \textbf{Speed}   & \textbf{$\Delta$V}    & \textbf{$\Delta$V}    & \textbf{Long.}  & \textbf{Lat.}   \\
 & \textbf{}          & \textbf{}        & \textbf{kms$^{-1}$}   & \textbf{kms$^{-1}$}   & \textbf{kms$^{-1}$} & \textbf{kms$^{-1}$} & \textbf{deg}    & \textbf{deg}    \\ \hline
1 & 2024 YR4       & 2040 JUL 24 & 0       & 0.0547  & 0.0547 & 0.0547 & 101.902  & -0.679 \\ \hline
2 & DSM @ 3.392 au & 2043 FEB 10 & 12.3925 & 12.3925 & 0      & 0.0547 & -111.135 & -1.293 \\ \hline
3 & DSM @ 3.606 au & 2045 MAY 24 & 11.0692 & 11.0694 & 0.001  & 0.0557 & -154.201 & -3.149 \\ \hline
4 & Earth          & 2047 DEC 23 & 13.234  & 0       & 0      & 0.0557 & 90.854   & -0.006 \\ \hline
\end{tabular}

\caption{Sample return segment of mission trajectory.}
\label{tab:DSMYR4_ret}
\end{table}

\begin{table}[]
\centering
\begin{tabular}{ccccccccc}
\hline
\textbf{} & \textbf{}          & \textbf{}     & \textbf{Arrival} & \textbf{Depart.} & \textbf{}          & \textbf{Cum.}      & \textbf{Eclip.} & \textbf{Eclip.} \\
\textbf{} & \textbf{Encounter} & \textbf{Time} & \textbf{Speed}   & \textbf{Speed}   & \textbf{$\Delta$V} & \textbf{$\Delta$V} & \textbf{Long.}  & \textbf{Lat.}   \\
\textbf{} & \textbf{}          & \textbf{}     & \textbf{$\si{km.s^{-1}}$}   & \textbf{$\si{km.s^{-1}}$}   & \textbf{$\si{km.s^{-1}}$}     & \textbf{$\si{km.s^{-1}}$}     & \textbf{$\si{deg}$}    & \textbf{$\si{deg}$}    \\ \hline
1 & Earth        & 2028 NOV 24 & 0       & 9       & 9      & 9      & 62.6   & 0    \\ \hline
2 & DSM @ 2.5 au & 2029 JUN 25 & 18.6055 & 18.508  & 0.3017 & 9.3017 & 175.3  & -3.3 \\ \hline
3 & DSM @ 1.1 au & 2032 NOV 17 & 35.7013 & 35.7048 & 0.01   & 9.3117 & 93.5   & -0.8 \\ \hline
4 & DSM @ 3.9 au & 2035 MAR 03 & 10.082  & 9.9318  & 0.2    & 9.5117 & -121.8 & -1.2 \\ \hline
5 & 2024 YR4     & 2037 JUL 29 & 0.3084  & 0       & 0.3084 & 9.8201 & -166.2 & -3.4 \\ \hline
\end{tabular}
\caption{Another possible rendezvous/loiter/sample return mission to 2024 YR$_4$, just involving Deep Space Manoeuvres and  no GAs.}
\label{tab:DSMYR4}
\end{table}

}

\hlreviewone{Note that the initial $\Delta$V at Earth of 9 $\si{km.s^{-1}}$, is defined here as the hyperbolic excess speed, V$_{\infty}$, equivalent to a launch characteristic energy, C$_3$ $= 81$ $\si{km^2s^{-2}}$. Referring to the NASA online Launch Vehicle Query portal \citep{NASALVQ}, a Falcon Heavy Expendable may launch a mass of 2115 $\si{kg}$ to this C$_3$, thus an OSIRIS-REx mission could indeed be undertaken to 2024 YR$_4$, as required.

For the OSIRIS-REx mission, it remained with Bennu for 2.5 years before returning home, after a further 2 or so years. For the equivalent mission to 2024 YR$_4$, it would take  around 8 years to eventually rendezvous with this target, which would happen around aphelion, where the spacecraft could remain with 2024 YR$_4$ for up to 4 years, allowing a further 7 years to return to Earth with negligible $\Delta$V requirement - refer Table \ref{tab:DSMYR4_ret}.

A similar mission profile to that articulated above is shown in Figure \ref{fig:Plotsrl2}, but with an extra Earth encounter instead of the final DSM. Although this seemingly has a higher in-flight $\Delta$V than that elaborated above, it should be noted that the NLP solver for the OITS tool took much longer to converge and may not have entirely found the optimal series of encounter times corresponding to minimum $\Delta$V. For this reason it is provided for completeness. Also refer to Table \ref{tab:DSMYR4} for detailed numerical data for this trajectory.

\begin{figure}[hbt!]
\centering
\includegraphics[width=1.0\textwidth]{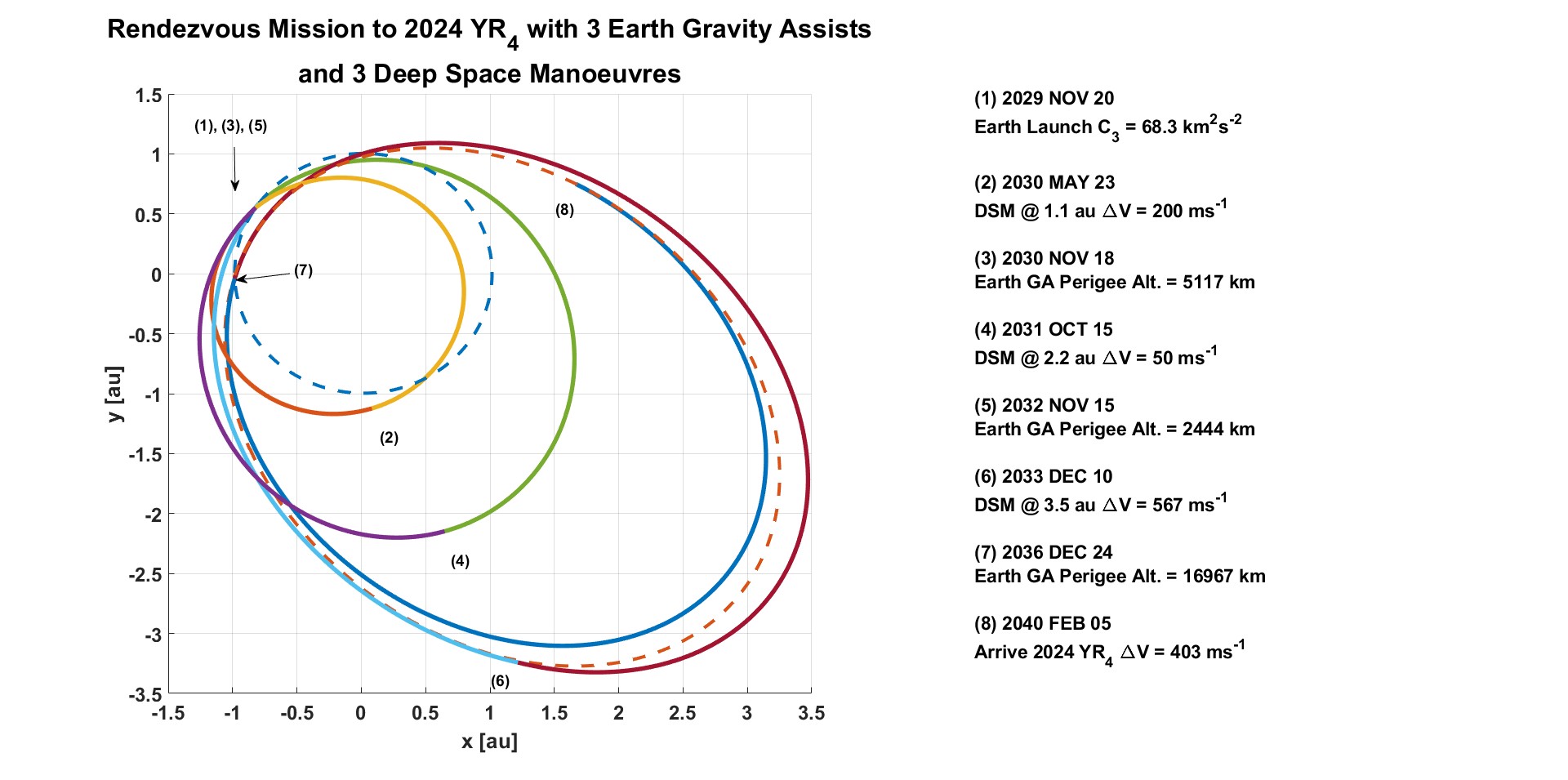}
\caption{Another trajectory for sample return with loiter, demonstrating the kinds of mission which might be flown to 2024 YR$_4$.}
\label{fig:Plotsrl2}
\end{figure}
}

\section{Rendezvous Without Lander and Without Gravity Assists}
\label{sec4}
Finally we investigate the possibility of \hlreviewone{\textbf{direct} (i.e. without any GAs)} rendezvous with 2024 YR$_4$, \hlreviewone{with no lander on-board and so NO sample return}. \hlreviewone{As already mentioned, a rendezvous} is defined as the mission scenario whereby the spacecraft has significant onboard propulsion enabling it to adjust its velocity vector so it may match velocity with the body in question, in this case 2024 YR$_4$. 

\begin{figure}[hbt!]
\centering
\includegraphics[width=1.0\textwidth]{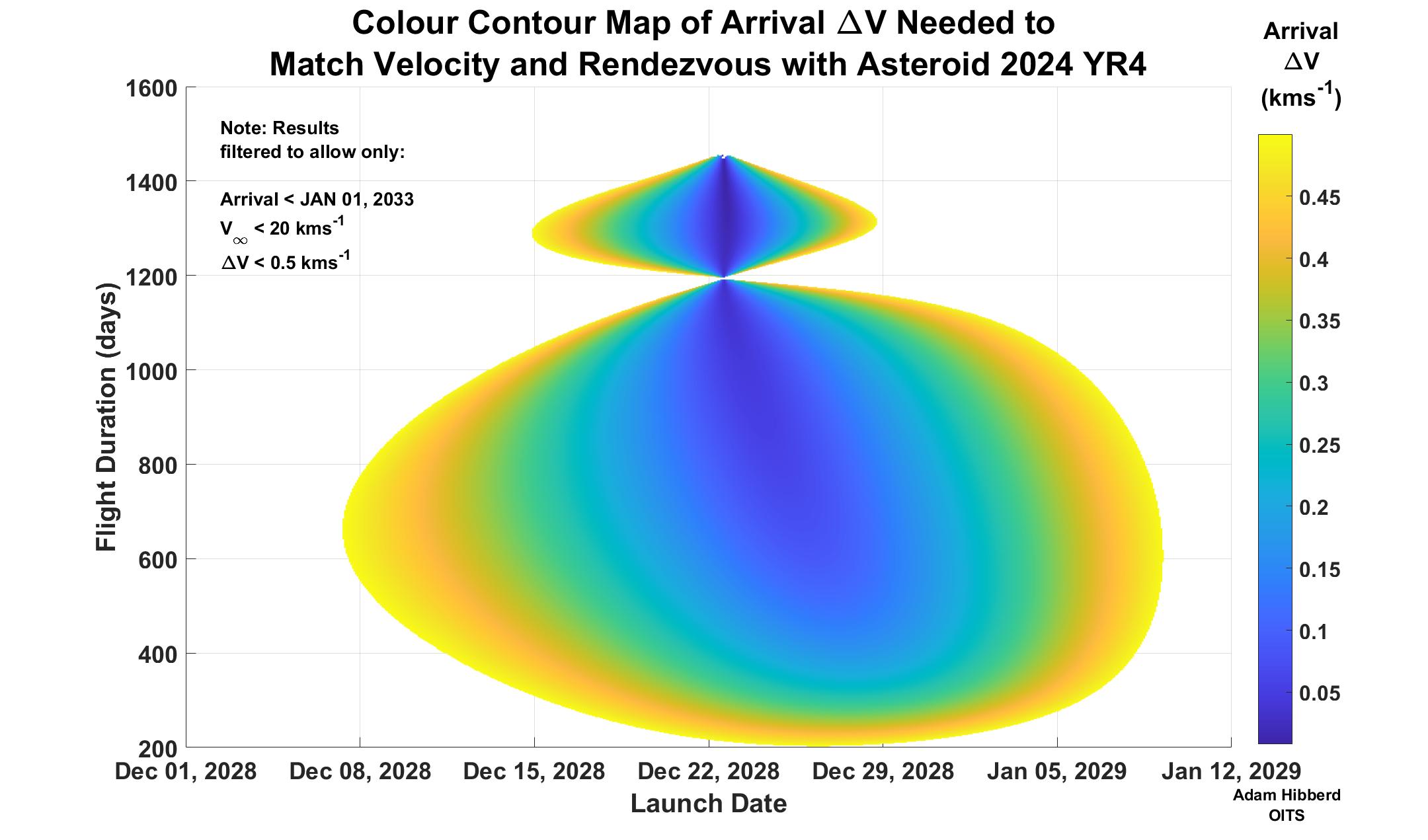}
\caption{Pork Chop plots indicating minimum arrival $\Delta$V needed by the spacecraft to match velocities with the object 2024 YR$_4$. A launch near the end of 2028 is favourable and similar launch windows will exist for the object's 2032 encounter.}
\label{fig:Plot4}
\end{figure}

\begin{figure}[hbt!]
\centering
\includegraphics[width=1.0\textwidth]{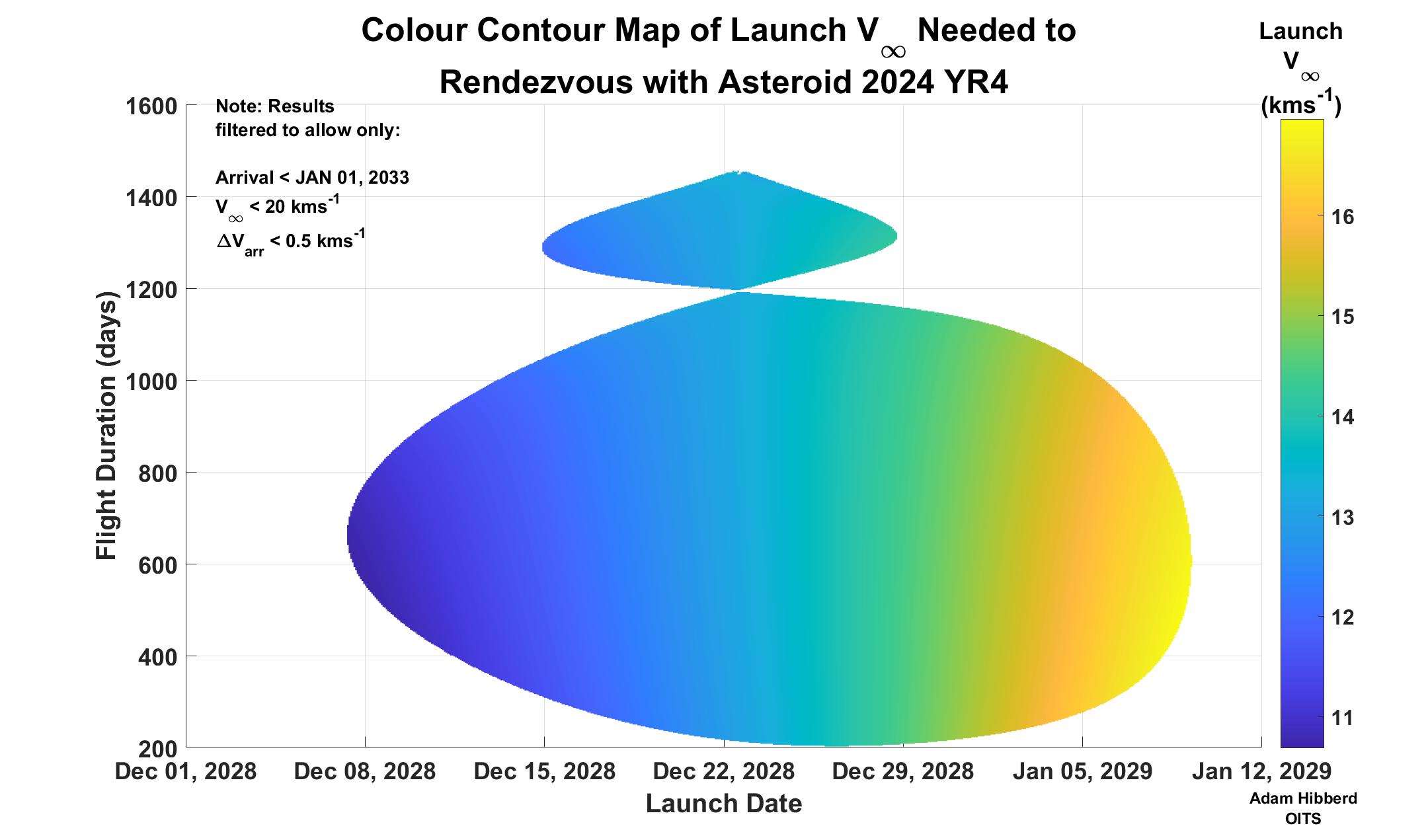}
\caption{Launch vehicle hyperbolic excess, $V_{\infty}$, required for the rendezvous missions indicated in Figure \ref{fig:Plot4}}
\label{fig:Plot5}

\end{figure}

If we refer to Figure \ref{fig:Plot4}, we find the possible mission launch dates and durations which will permit an impulsive change in velocity upon arrival ($\Delta$V) to be less than 0.5 \si{km.s^{-1}} which in turn enables a probe to remain with 2024 YR$_4$ in its path around the Sun. The corresponding level of $V_{\infty}$ at launch to permit this low arrival $\Delta$V  (Figure \ref{fig:Plot5}), is considerably higher than both flyby and sample return missions, since the launch occurs near to the descending node of 2024 YR$_4$, which in turn allows intercept at orbital radii much closer to the target's aphelion, the point where the object is travelling slowest.

As a reference let us consider the NASA spacecraft to Pluto launched in 2006, New Horizons, a mission design which has been demonstrated and proven to be successful, and therefore a replicable mission which is perfectly within the capabilities of humanity. Although clearly NOT a rendezvous mission, yet it had a considerable in-flight $\Delta$V performance of $\sim{0.3}$ \si{km.s^{-1}} \citep{NEWHORIZONS}, and as can be seen from Figure \ref{fig:Plot4} a value easily within the performance envelope required by a rendezvous of 2024 YR${_4}$.

New Horizons had a V$_{\infty}$ capability of $\sim{13.0}$ \si{km.s^{-1}} \citep{weaver2008new}, which is about the same as required for a rendezvous mission to 2024 YR$_{4}$ (ref. Figure \ref{fig:Plot5}). Moreover the arrival $\Delta$V needed by any putative spacecraft can be maintained well below the capability of New Horizon's propulsion system. 

\begin{figure}[hbt!]
\centering
\includegraphics[width=1.0\textwidth]{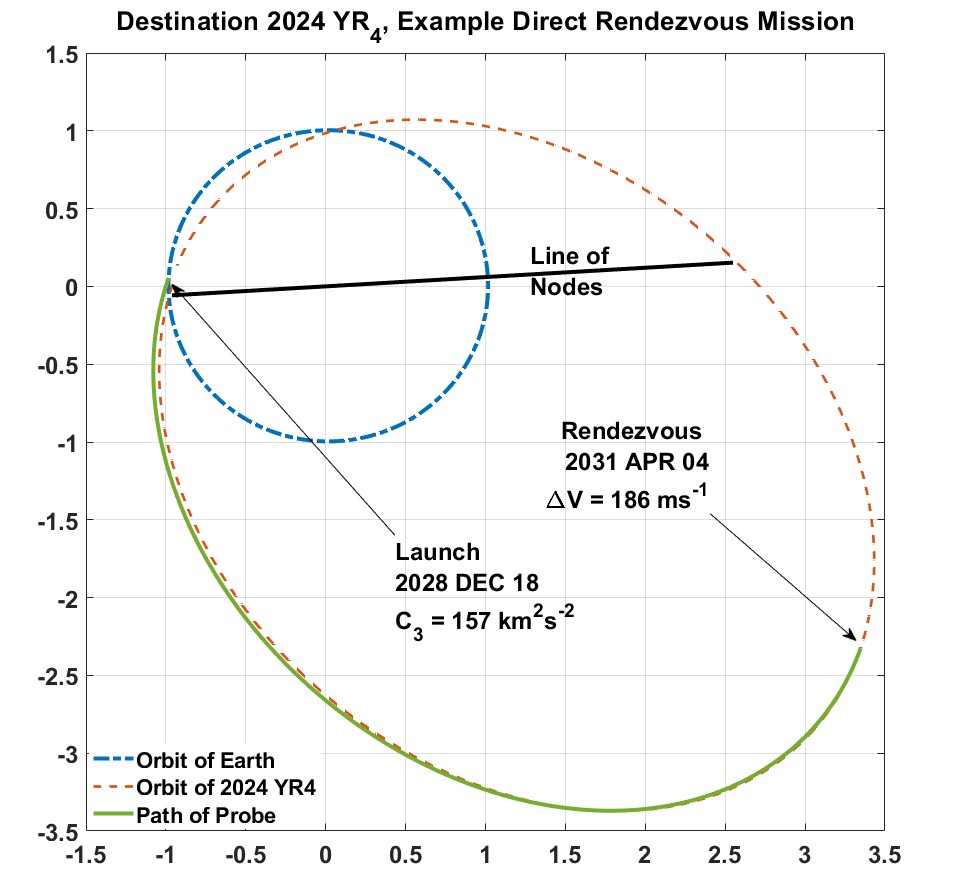}
\caption{Example direct rendezvous trajectory for a mission to 2024 YR$_4$, revealing that a New Horizons style spacecraft would be feasible to this object.}
\label{fig:Plotrt}
\end{figure}

\hlreviewone{An example of such a mission is depicted in Figure \ref{fig:Plotrt}, with launch in December 2028 and arrival in April 2031.}


\section{Discussion}
\label{sec5}
The research conducted here reveals that the orbit of PHO 2024 YR$_4$ is quite serendipitous in that many launch windows are available to us should it be desirable or even necessary for us to send a mission.

Whether the mission should be a flyby, sample return, or rendezvous is dependent on what would actually need to be achieved by such a mission. For instance a flyby, with relative arrival velocities of $\sim{13.5}$ \si{km.s^{-1}} would be ideal if a mission to deflect it was deemed necessary.  Such a mission could deploy an impactor to be targeted on the object in a similar manner the DART spacecraft successfully deflected asteroid Dimorphos, a minor asteroid in a binary asteroid system.

Alternatively if an impactor mission was conducted, and in addition an orbital transfer trajectory of 1 au was selected (refer section \ref{sec3}) this flyby could additionally return to Earth and achieve a sample return, with the return Earth hyperbolic excess virtually negligible.

Rendezvous missions are also available (Section \ref{sec4}), and could involve determining the mass and moreover characterizing the gravitational field of 2024 YR$_4$, by analysing the motion of the probe relative to it and how it is influenced by its gravity. Alternatively a VLBI beacon could be placed on the object to allow accurate determination of its position and velocity over the course of time.

\hlreviewone{Furthermore a rendezvous/loiter/sample return mission with on-board lander of the kind exemplified by the OSIRIS REx mission to asteroid Bennu, is also feasible to 2024 YR$_4$ with innumerable rewards in terms of scientific return (Section \ref{secV}). }

In short, many, many mission architectures are foreseeable for 2024 YR$_4$, offering an 'opportunity rich environment', and places humanity, and more specifically space mission designers, in an ideal position for realising a mission of choice to this body, and achieving a hitherto unattainable scientific return.   

\section{Conclusion}
\label{sec6}
The research in this paper employed OITS to examine the general feasibility of missions to 2024 YR$_{4}$, and many such opportunities and launch windows were discovered. Specifically for flyby missions, launches in 2028 or 2032 are readily achieved, with associated negligible launch  $C_3$ magnitudes ($\sim{0.0}$ \si{km^{2}s^{-2}}) lasting almost a year for both cases.

Both sample return missions and rendezvous missions were also found to be feasible. For instance a rendezvous mission launched in late 2028/early 2029 could easily be conducted using a NASA New Horizons type craft, which has a $\Delta$V capability quite sufficient to achieve the necessary rendezvous. 

\bibliography{Missions_to_2024_YR4}

\end{document}